\documentclass[sigconf]{acmart}
\AtBeginDocument{%
  \providecommand\BibTeX{{%
    \normalfont B\kern-0.5em{\scshape i\kern-0.25em b}\kern-0.8em\TeX}}}

\def \toolname {CocciEvolve}

\definecolor{red}{HTML}{9B0000}
\definecolor{lightred}{HTML}{FF5131}
\definecolor{green}{HTML}{006400}
\definecolor{lightgreen}{HTML}{9CFF57}
\definecolor{purple}{HTML}{7200CA}
\definecolor{verylightgrey}{HTML}{F1F1F1}

\definecolor{diffstart}{named}{blue}
\definecolor{diffincl}{named}{green}
\definecolor{diffrem}{named}{red}

\newcommand{\extrabold}{\bfseries}

\usepackage{multirow}
\usepackage{listings}
  \lstdefinelanguage{diff}{
	basicstyle=\ttfamily\extrabold\scriptsize,
	morecomment=[f][\color{diffstart}]{@},
	morecomment=[f][\color{diffincl}]{+},
	morecomment=[f][\color{diffrem}]{-},
        keepspaces=true,
	identifierstyle=\color{black},
  }

\usepackage[shortlabels]{enumitem}

\copyrightyear{2020} 
\acmYear{2020} 
\setcopyright{acmcopyright}
\acmConference[ICPC '20]{28th International Conference on Program Comprehension}{October 5--6, 2020}{Seoul, Republic of Korea}
\acmBooktitle{28th International Conference on Program Comprehension (ICPC '20), October 5--6, 2020, Seoul, Republic of Korea}
\acmPrice{15.00}
\acmDOI{10.1145/3387904.3389285}
\acmISBN{978-1-4503-7958-8/20/05}

\begin{document}
\title{Automatic Android Deprecated-API Usage Update \protect\\by Learning from Single Updated Example}

\author{Stefanus A. Haryono$^*$, Ferdian Thung$^*$, Hong Jin Kang$^*$, Lucas Serrano$^{\dagger}$, Gilles Muller$^{\ddagger}$,\ \ \ \ \ \ \ \ \ \ \ \ \ \ \ \ \ \ \  Julia Lawall$^{\ddagger}$, David Lo$^*$, Lingxiao Jiang$^*$}
\affiliation{%
  \institution{*School of Information Systems, Singapore Management University, Singapore}
}
\email{{stefanusah,ferdianthung,hjkang.2018,davidlo,lxjiang}@smu.edu.sg}
\affiliation{%
  \institution{$^{\dagger}$Sorbonne University/Inria/LIP6, France}
}
\email{Lucas.Serrano@lip6.fr}
\affiliation{%
  \institution{$^{\ddagger}$Inria, France}
}
\email{{Gilles.Muller,Julia.Lawall}@inria.fr}

\begin{abstract}
Due to the deprecation of APIs in the Android operating system, developers have to update usages of the APIs to ensure that their applications work for both the past and current versions of Android.
Such updates may be widespread, non-trivial, and time-consuming.
Therefore, automation of such updates will be of great benefit to developers. 
AppEvolve, which is the state-of-the-art tool for automating such updates, relies on having before- and after-update examples to learn from. In this work, we propose an approach named \toolname\ that performs such updates using only a single after-update example. 
\toolname\ learns edits by extracting the relevant update to a block of code from an after-update example. From preliminary experiments, we find that \toolname\ can successfully perform 96 out of 112 updates, with a success rate of 85\%.
\end{abstract}

\maketitle
\renewcommand{\shortauthors}{Haryono, et al.}

\section{Introduction}
When an Android API is deprecated, apps using the API should update their usages of the API to ensure that they still work in the current and future versions of Android.
For these updates, developers need to learn the new API(s) that should replace the deprecated API, while maintaining backward compatibility with older versions to address Android fragmentation~\cite{he2018understanding,li2018cid}. Moreover, the deprecated API may be used in multiple locations in a codebase. 
Thus, manually updating the deprecated API may be cumbersome and time consuming. 

To help developers in updating usages of deprecated APIs with their replacement APIs, Fazzini et al.~\cite{fazzini2019automated} proposed AppEvolve to automate the update task. AppEvolve learns to transform applications that use a deprecated API by learning from before- and after-update examples in GitHub.
These updates add usages of replacement APIs around usages of the deprecated API along with conditional checks of Android versions in code. AppEvolve learns a generic patch from such examples and applies the transformation from each generic patch in a certain order. 

Recently, Thung et al.~\cite{thung2020automated} reported that, in order for AppEvolve to perform a successful update, the target code requiring update has to be written  syntactically similar to the before- and after-update example. They demonstrated that AppEvolve's performance can be improved significantly if the app code is {\em manually} rewritten to have syntactic similarities to the before- and after-update example.

\begin{figure}[t]
	\centering
	\scriptsize{
\begin{lstlisting}[language=java,numbers=none,sensitive=true,columns=flexible,basicstyle=\ttfamily]
if (android.os.Build.VERSION.SDK_INT >= android.os.Build.VERSION_CODES.M) {
    hour = picker.getHour();
} else {
    hour = picker.getCurrentHour();
}
\end{lstlisting}
		\caption{An example of an after-update for  {\tt getCurrentHour} deprecated API}\label{fig:after_update_example}
	}
\end{figure}

In this work, we propose \toolname{}. \toolname{} outperforms AppEvolve in the following aspects:
\begin{enumerate}[nosep,leftmargin=*]
    \item \toolname\ eliminates the shortcoming of AppEvolve by normalizing both the after-update example and the target app code to update. In this way, \toolname\ ensures that both of them are written similarly, thereby preventing unsuccessful updates caused by minor differences in the way the code is written.
   
    \item \toolname\ requires only a single after-update example for learning how to update an app that uses a deprecated API. 
    Consider an after-update example in Figure~\ref{fig:after_update_example}. The {\tt if} and {\tt else} blocks correspond to the code using the replacement and deprecated API methods, respectively. The code in the {\tt if} block runs only on versions of Android after the deprecation. Thus, the code in {\tt else} block can be considered as the code that uses the deprecated method before the update.  
    \item Transformations made by \toolname\ are expressed in the form of a semantic patch by leveraging Coccinelle~\cite{lawall2018coccinelle}. Semantic patch has a syntax similar to a \textit{diff} that is familiar to software developers. Therefore, the modifications process are more readable and understandable to developers.
\end{enumerate} 

\noindent
The contributions of our work are:
\begin{itemize}[nosep,leftmargin=*]
    \item We propose \toolname, an approach for updating Android deprecated-API usages using only a single after-update example. 
    \item We perform {\em automatic} code normalization of both the update example and the target code to be updated, addressing the challenge of updating code that is semantically equivalent but syntactically different, which was a limitation of prior work.
    \item 
We have evaluated \toolname\ with a dataset of 112 target files to update that we obtained from Github. The 112 files use 10 deprecated APIs used in the original evaluation of AppEvolve. 
We show that \toolname\ can successfully update 96 target files. 

\end{itemize}

\noindent
The remainder of this paper is structured as follows. Section~\ref{sec:prelim} provides some preliminaries. Section~\ref{sec:approach} details our proposed approach. 
Section~\ref{sec:exp} describes our preliminary experiments and results. 
Section~\ref{sec:related} presents related work. Finally, we conclude in Section~\ref{sec:conclusion}.

\section{Preliminaries}\label{sec:prelim}
{\bf AppEvolve.}
AppEvolve is the state-of-the-art tool automating API-usage updates for deprecated Android APIs. As input, it takes an app to update and a mapping from a deprecated method to its replacement method(s). It has four phases: {\em API-Usage Analysis}, {\em Update Example Search}, {\em Update Example Analysis}, and {\em API-Usage Update}. 

In the {\em API-Usage Analysis} phase, AppEvolve finds uses of the deprecated method inside the app. 
In the {\em Update Example Search} phase, AppEvolve searches GitHub for apps that use both the deprecated and replacement methods in the same file. 
For each of these apps, AppEvolve searches the app commit history for a change that adds the replacement method(s) without removing the deprecated method. These changes are used to learn how to update deprecated method usages. In the {\em Update Example Analysis} phase, AppEvolve produces a generic patch from each example. AppEvolve then computes the
common core from the produced generic patches. The common core is the longest subsequence of edits across the patches. In the {\em API-Usage Update} phase, AppEvolve applies generic patches in ascending order of their distance to the common core. To apply a generic patch, AppEvolve tries to match context variables to variables in the app. If matches are found, AppEvolve applies edits in the generic patch and returns the updated app if the edits are successful.

\vspace{0.2cm}\noindent{\bf Coccinelle4J.}
Coccinelle4J~\cite{kang2019semantic} is a recent Java port of Coccinelle, 
which is a program matching and transformation tool \cite{lawall2018coccinelle,padioleau2008documenting}.
Given the source code of a program and a {\em semantic patch} describing the desired transformations, Coccinelle4J transforms the parts of the source code that match the semantic patch. 

Written in the Semantic Patch Language (SmPL), 
a semantic patch has two parts: (1) context information, including the declaration of metavariables; and (2) changes to be made to the source code. 
A metavariable can match program elements of the type indicated in its declaration. Modifications are expressed using fragments of Java as follows: (1) code that should be removed by annotating the start of the lines with $-$; and (2) code that should be added by annotating the start of the lines with $+$. Unannotated lines add context to the semantic patch. 
Figure~\ref{fig:semantic_patch_example} showed an example of a semantic patch. The line surrounded by {\tt@@} declares the metavariable.

\begin{figure}[h]
	\centering
	\scriptsize{
\begin{lstlisting}[language=diff,numbers=none]
@@
expression timepicker;
@@
- timepicker.getCurrentHour();
+ timepicker.getHour();
\end{lstlisting}
		\caption{A simplified example of a semantic patch that transforms uses of the deprecated method, {\tt getCurrentHour} }\label{fig:semantic_patch_example}
	}
\end{figure}

\section{Approach}\label{sec:approach}

\begin{figure}[t]
	\centering
	\includegraphics[width=0.95\linewidth]{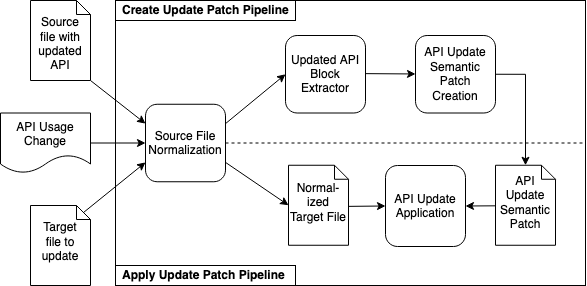}
	\caption{System overview of CocciEvolve}
	\label{fig:framework}
\end{figure}

An overview of our proposed system and the relevant pipelines are shown in Figure~\ref{fig:framework}. \toolname\ is built on three main components: (1) Source file normalization, (2) Updated API block detection, and (3) API-update semantic patch creation. These components are the building blocks of the \toolname\ pipelines: a pipeline to create the update semantic patch, and a pipeline to apply the update to a target file. 
The create update patch pipeline takes as inputs the API usage change, and a source file containing updated API call. 
The apply update pipeline takes as inputs the API usage change, target source file, and the update semantic patch file. In the following subsections, we will explain in detail each of the system components.



\subsection{Source File Normalization}
Different software developers may have different programming styles, thus, semantically-equivalent code may be expressed in different syntactic forms. As a result, equivalent usages of one API may vary in its expression
. Figure~\ref{fig:syntax_different_same_semantic} shows an example of such cases in which \texttt{getCurrentHour()} is expressed differently. 
Therefore, it is necessary to perform source code normalization. 

\begin{figure}[t]
	\centering
	\scriptsize{
\begin{lstlisting}[language=Java,numbers=none,sensitive=true,columns=flexible,basicstyle=\ttfamily]
if (timePicker.getCurrentHour() > 11)
    itsNoon();
    
int currentHour = timePicker.getCurrentHour();
if (10 < currentHour)
    itsNoon();
\end{lstlisting}
		\caption{Two fragments of semantically-equivalent code expressed differently}\label{fig:syntax_different_same_semantic}
	}
\end{figure}

Focusing on source code that is related to the API usage, we perform the following code normalization steps:
\begin{itemize}[nosep,leftmargin=*]
    \item An API call contained in a compound expression or statement (e.g. if, loop, return, etc) is extracted into a variable assignment. 
    \item Arguments of an API call are extracted into variable assignments.
    \item The object receiving an API call is extracted into a variable assignment.
\end{itemize}
Two components are responsible for normalization in \toolname{}: 

\subsubsection{Statement Extractor}
For calls to API methods that return a value, the Statement Extractor extracts API calls that are part of compound expressions or statements. 
Before each of such a compound expression or statement, a new simple statement is inserted, 
which initializes a new temporary variable with the return value of the API call.
This extraction is performed by using Coccinelle4J, with a semantic patch that is built from the type signature of the input API. 
An example of statement extraction can be seen in Figure~\ref{fig:statement_extraction}.

\begin{figure}[t]
	\centering
	\scriptsize{
\begin{lstlisting}[language=diff,numbers=none]
final String tmDevice;
+ String tempFunctionReturnValue;
+ tempFunctionReturnValue = tm.getDeviceId();
- tmDevice = "" + tm.getDeviceId();
+ tmDevice = "" + tempFunctionReturnValue;
\end{lstlisting}
		\caption{Example of statement extraction for {\tt getDeviceId} API call in a compound expression.}\label{fig:statement_extraction}
	}
\end{figure}

\subsubsection{Variable Extractor}
The Variable Extractor is used to extract the arguments and object that the API call is invoked on. 
These arguments and object are assigned to temporary variables. 
As before, this extraction is done through Coccinelle4J using a semantic patch.
An example of this extraction is shown in Figure~\ref{fig:variable_extraction}.

\begin{figure}[t]
	\centering
	\scriptsize{
\begin{lstlisting}[language=diff,numbers=none]
+ Context paramVar0 = context;
+ int paramVar1 = android.R.style.TextAppearance_Large;
+ TextView classNameVar = tvTitle;
- tvTitle.setTextAppearance(context, 
-   android.R.style.TextAppearance_Large);
+ classNameVar.setTextAppearance(paramVar0, paramVar1);
\end{lstlisting}
		\caption{Example of variable extraction for call to method {\tt setTextAppearance}}\label{fig:variable_extraction}
	}
\end{figure}

\subsection{Updated API Block Extractor}
The Updated API Block Extractor identifies the relevant block of code containing the updated API call. 
This block is extracted and used as the source of the update patch. 
In order to prevent false positives due to irrelevant code blocks, we use the following rules to identify a valid updated API block:
\begin{enumerate}[nosep,leftmargin=*]
    \item A valid updated block contains the updated API call in the {\tt if} branch and the old API call in the {\tt else} branch or vice versa. 
    \item A valid updated block will have an Android version check as the {\tt if} condition.
\end{enumerate}
These classification rules are implemented in two components: 

\subsubsection{Version Statement Normalization}
One of the important criteria for a valid updated block is the presence of
an if statement that checks for the Android version. 
However, this check may take different forms in different projects (e.g. the Android version may first be assigned to a local variable). 
To alleviate this problem, \toolname{} normalizes the conditions involving the Android version. 
This normalization is done automatically by first detecting any assignment of an Android version constant to a variable, and then replacing usage of the local variable in a condition with the relevant Android version constant. 
An example can be seen in Figure~\ref{fig:version_normalization}.

\begin{figure}[h]
	\centering
	\scriptsize{
\begin{lstlisting}[language=diff,numbers=none]
int currentBuildVersion = Build.VERSION.SDK_INT;
int marshmallowVersion = Build.VERSION_CODES.M;
- if (currentBuildVersion >= marshmallowVersion) {
+ if (Build.VERSION.SDK_INT >= Build.VERSION_CODES.M) {
    timePicker.setHour(1);
}
\end{lstlisting}
		\caption{Sample Android version statement normalization}\label{fig:version_normalization}
	}
\end{figure}

\subsubsection{Update Block Extractor}
The Update Block Extractor takes as an input the source file that has been normalized, then it detects a valid block based on the aforementioned criteria and extracts it from the file. This block is input to the API Update Semantic Patch Creation component.

\subsection{API Update Semantic Patch Creation}
Using the normalized update block as an input, this component creates a Coccinelle4J semantic patch that can be used to update a normalized target file. 
This component will replace variables and expressions with metavariables. 
Metavariables will bind to program elements in the input code passed into Coccinelle4J. 
The patch works by detecting the location of the old API call and then adding the new code which consists of a surrounding if block, the updated API call, and new variables introduced by the new API. 

To increase the robustness of the system, for APIs that return a value, two different update rules are created. 
One rule is for cases where the return value is assigned into a variable, while the other is for cases where the return value is not used. 
An example of the update semantic patch can be seen in Figure~\ref{fig:cocci_example}.

\begin{figure}[h]
	\centering
	\scriptsize{
\begin{lstlisting}[language=diff,numbers=none]
@bottomupper@
expression exp0;
identifier classIden;
@@
TimePicker classIden = exp0;
...
+ if (Build.VERSION.SDK_INT >= Build.VERSION_CODES.M) {
+ classIden.getMinute();
+ } else {
 classIden.getCurrentMinute();
+ }

@bottomupper_assignment@
expression exp0;
identifier classIden;
@@
TimePicker classIden = exp0;
...
+ if (Build.VERSION.SDK_INT >= Build.VERSION_CODES.M) {
+ tempFunctionReturnValue = classIden.getMinute();
+ } else {
tempFunctionReturnValue = classIden.getCurrentMinute();
+ }
\end{lstlisting}
		\caption{Example of update patch for {\tt getCurrentMinute} API}\label{fig:cocci_example}
	}
\end{figure}

\section{Experiments}\label{sec:exp}

\subsection{Dataset}
To assess the performance of \toolname\ for practical usage, 
we use a dataset of real-world Android projects retrieved from public Github repositories. 
For this purpose, we use AUSearch~\cite{asyrofi2020ausearch}, a tool for searching Github repositories, to find Android API usage. 
For the update semantic patch creation, we use the existing after-update examples provided in the AppEvolve replication package. 
For each API, only a single after-update example is used.
We obtained a total of 112 target source files from Github for 
the 10 most commonly used APIs that were used in the original evaluation of AppEvolve. These target files are disjoint from the target files used by AppEvolve and thus are used to evaluate AppEvolve's generalizability in updating other target files.
Detailed statistics for this dataset can be seen in Table~\ref{table:data_statistic}. 
This dataset is published with our replication package.\footnote{\label{refnote}\url{https://sites.google.com/view/cocci-evolve/}}

\begin{table}[t]
\caption{Number of targets in our evaluation dataset }
\begin{center}
\begin{tabular}{ |p{20em}|p{4.5em}| }
\hline
 \textbf{API Description} & \textbf{\# Targets}
\\ \hline

getAllNetworkInfo() & 8
\\ \hline
getCurrentHour() & 9
\\ \hline
getCurrentMinute() & 12
\\ \hline
setCurrentHour(Integer) & 12
\\ \hline
setCurrentMinute(Integer) & 10
\\ \hline
setTextAppearance(...) & 12
\\ \hline
release() & 14
\\ \hline
getDeviceId() & 12
\\ \hline
requestAudioFocus(...) & 8
\\ \hline
saveLayer(...) & 15
\\ \hline

\end{tabular}
\end{center}
\label{table:data_statistic}
\end{table}

\subsection{Experimental Settings}
Our experiments are done by comparing the performance of \toolname\ against AppEvolve based on the number of applicable updates produced. 
To generate
the update patch, we utilize a single update example for each API from the available AppEvolve examples.

The target files for the experiments are the public Android project dataset that have been collected from Github through the use of AUSearch~\cite{asyrofi2020ausearch}. \toolname\ is applied to every target file using the relevant API update patch that was created. 
For experiments involving AppEvolve, we need to configure each target project as an Eclipse project and create an additional XML file that contains the deprecated API description and their locations in the file. 
Due to this limitation, our experiments on AppEvolve are focused on first instance of each API call for each target project. 

\subsection{Results}
In our experiments, \toolname\ attains a better performance compared to AppEvolve. 
For most APIs, \toolname\ achieves a near perfect result. 
We ask a software engineer with three years experience in Android, who was not part of this project, to validate the correctness of the update by verifying that there are no semantic changes in the update. Our experimental results are also included in our replication package.

In most cases, AppEvolve does not produce any code update. Thung et al.~\cite{thung2020automated} note that AppEvolve requires some manual code refactoring and modifications to be able to perform the automated update. Table~\ref{result_statistic} shows the statistics of our evaluation.

\begin{table}[t]
  \centering
  \caption{Statistics of updating target files per API}
  \label{result_statistic}
  \begin{tabular}{|l|c|c|c|c|}
 
 \hline
\multicolumn{1}{|c|}{\multirow{2}{*}{\textbf{API}}}  & \multicolumn{2}{c|}{\textbf{CocciEvolve}} & \multicolumn{2}{c|}{\textbf{AppEvolve}} \\ 
\cline{2-5} 
\multicolumn{1}{|c|}{}                     & 
\textbf{Success}           & \textbf{Fail}          & \textbf{Success}         & \textbf{Fail}         \\ \hline

 getAllNetworkInfo()  & 0 & 8 & 0 & 8 
\\ \hline
getCurrentHour()  & 9 & 0 & 1 & 8
\\ \hline
getCurrentMinute() & 12 & 0 & 1 & 11
\\ \hline
setCurrentHour(Integer)  & 12 & 0 & 10 & 2
\\ \hline
setCurrentMinute(Integer)  & 10 & 0 & 6 & 4
\\ \hline
setTextAppearance(...) & 12 & 0 & 1 & 11
\\ \hline
release() & 14 & 0 & 0 & 14
\\ \hline
getDeviceId() & 12 & 0 & 1 & 11
\\ \hline
saveLayer(...)  & 15 & 0 & 0 & 15
\\ \hline
requestAudioFocus(...)  & 0 & 8 & 0 & 8
\\ \hline
\textbf{Total}  & \textbf{96} & \textbf{16} & \textbf{20} & \textbf{92}
\\ \hline 
  \end{tabular}
  
\end{table}

Based on the experiments result, we can see that \toolname\ mainly failed for two APIs: {\tt getAllNetworkInfo()} and {\tt requestAu}-{\tt dioFocus(...)}.  
Updating these two APIs requires the creation of new objects for arguments to the replacement APIs.
These objects are frequently created outside of the updated API block, 
and will require a data flow analysis to detect and construct the update correctly. The current version of \toolname{} does not support sophisticated data-flow analysis.

Compared to AppEvolve, \toolname{} has several advantages:

\begin{itemize}[nosep,leftmargin=*]
    \item \toolname\ does not need extensive setup or configuration;
    \item \toolname\ is capable of updating multiple API calls in the same file without additional configuration;
    \item \toolname\ provides an easily readable and understandable semantic patch as a by-product;
    \item \toolname\ only needs a single updated example.
\end{itemize}

\section{Related Work}\label{sec:related}
There are many studies on API deprecation~\cite{kapur2010refactoring,zhou2016api,brito2016developers,sawant2018features,li2018characterising,CiD,ACRyL,PIVOT,ELEGANT,understandingfic,tamingandroid}. 
Kapur et al.~\cite{kapur2010refactoring} discovered that  APIs can be removed from a library without warning. 
Zhou and Walker~\cite{zhou2016api} proposed a tool to mark deprecated API usages in StackOverflow posts. 
Some studies propose 
approaches~\cite{CiD,ACRyL, PIVOT,ELEGANT,understandingfic,tamingandroid} to detect API compatibility issues.
Brito et al.~\cite{brito2016developers} showed that not all APIs are annotated with replacement messages.
Sawant et al.~\cite{sawant2018features} found 12 reasons for deprecation. 
Li et al.~\cite{li2018characterising} 
characterized Android APIs and 
found inconsistencies between their annotation and documentation. Unlike these studies, our work aims to automatically update usages of deprecated Android APIs.

There are many studies on program transformations inference~\cite{Meng:2013:LLA:2486788.2486855,Rolim:2017:LSP:3097368.3097417,Rolim:arxiv,jiang2019inferring,fazzini2019automated}.
LASE~\cite{Meng:2013:LLA:2486788.2486855} creates edit scripts by finding common changes from a set of change examples.
REFAZER~\cite{Rolim:2017:LSP:3097368.3097417} employs a programming-by-example methodology to infer transformations from a set of change examples. REVISAR~\cite{Rolim:arxiv} finds common Java edit patterns from code repositories by clustering the edit patterns. Jiang et al.~\cite{jiang2019inferring} proposed GenPat, which builds source code hypergraphs to infer transformations. 
Fazzini et al.~\cite{fazzini2019automated} proposed AppEvolve to transform app with Android deprecated-API usages into ones that are backward compatible.
Our work shares the same goal. However, while AppEvolve learns from a before- and after- update example, our approach requires only one after-update example.

\section{Conclusion and Future Work}\label{sec:conclusion}

In this work, we propose \toolname{}, which can learn an Android API update from a single after-update example. 
\toolname\ performs code normalization to standardize the code and writes the update in the Semantic Patch Language (SmPL) to make the transformation transparent and readable to developers.
Our experiments on 112 target files on 10 deprecated Android APIs show that \toolname\ successfully updates usages in 96 target files. On the other hand, AppEvolve can only update deprecated-API usages in 20 of these target files. For future work, we plan to perform code slicing to find code relevant to the API update, including code beyond method boundaries. We also plan to perform code denormalization to restore the original coding style.

\noindent{\bf Acknowledgement.} This research is supported by the Singapore NRF (award number: NRF2016-NRF-ANR003) and the ANR ITrans project.

\bibliography{references}
\bibliographystyle{ACM-Reference-Format}

\end{document}